\documentclass[onecolumn]{article}
\input{epsf.sty}

\title{Depolarization of multiple scattered light in atmospheres due to anisotropy of small grains and molecules.
II. The problems with sources}

\author{N. A. Silant'ev\thanks{E-mail:nsilant@bk.ru}\,\,, G. A. Alekseeva,\, V. V. Novikov
\medskip\\ Central Astronomical Observatory at Pulkovo of Russian Academy of Sciences,\\ 196140,
Saint-Petersburg, Pulkovskoe shosse 65, Russia}

\begin{document}

\maketitle

\begin{abstract}
In the previous paper we considered two classic problems - the diffuse reflection of the light beam from semi-infinite atmosphere, and the Milne problem. For both problems we used the technique of invariance principle.
 In this paper we consider the solution of the problem when in semi-infinite atmosphere the sources of unpolarized
 radiation are exist. Here we used the technique of the Green matrices. We consider only continuum radiation.

\end{abstract}

\section{Introduction}

Real atmospheres of stars and planets, and optically thick accretion disks consist of anisotropic particles.
Usually these particles (dust grains and molecules) are small compared to the wavelengths of radiation.
To consider the process of multiple light scattering in such atmospheres we have to remember the features of
single scattering.

The incident electromagnetic wave ${\bf E}(\omega)$ induces  dipole moment in the scattering particle,
$p_i(\omega)=\beta_{ij}(\omega)E_j(\omega)$. This dipole moment gives rise to scattered radiation.
The polarizability tensor $\beta_{ij}(\omega)$ characterizes the structure of the dust grains and molecules.
Anisotropic particle with axial symmetry is characterized by two polarizabilities - along the symmetry axis
$\beta_{\parallel}(\omega)$, and in transverse direction - $\beta_{\perp}(\omega)$. For isotropic particles
(electron, spherical dust grains) polarizabilities $\beta_{\parallel}=\beta_{\perp}$.
It is known (see, for example, Dolginov et al. 1995, Gnedin and Silant'ev 1997) that for $\beta_{\parallel}=\beta_{\perp}$
the polarization of scattered radiation is greater than that for anisotropic particles with $\beta_{\parallel}\neq
\beta_{\perp}$.

The anisotropic scattering particles due to chaotic thermal motions are freely oriented in an atmosphere.
The radiative transfer equation for this case firstly was derived by Chandrasekhar (1960). 
The average over all orientations of particles results to the additional term, depending on anisotropy
of particles. This term depends integrally on the intensity $I(\tau,\mu)$ of radiation, and gives rise to additional
isotropic unpolarized part of scattered radiation. For this reason, the depolarizing effect of anisotropy of particles mostly reveals by consideration of axially symmetric problems. One of such problem is the multiple scattering of light in an
semi-infinite atmosphere with sources of unpolarized radiation.

In such problem the radiation is described by two intensities - $I_l(\tau,\mu)$ and $I_r(\tau,\mu)$.
Here $\tau $ is the optical depth below the surface of semi-infinite plane-parallel atmosphere,
$\mu =\cos\vartheta$ with $\vartheta$ being the angle between the outer normal ${\bf N}$ to the surface and the direction
of light propagation ${\bf n}$.
 The intensity $I_l$ describes the light linearly polarized in the plane (${\bf nN}$), and $I_r$ has polarization
perpendicular to this plane. The total intensity $I=I_l+I_r$, and the Stokes parameter $Q=I_l-I_r$. The Stokes
parameter $U\equiv 0$. The degree of linear polarization is equal to $p=|I_l-I_r|/(I_l+I_r)$.
Circularly polarized light is described by the separate equation. We do not consider this equation.

A detailed consideration of problems without parameter $\overline{b}_2$ is presented in many papers (see,
for example, Chandrasekhar 1960; Horak \& Chandrasekhar 1961; Lenoble 1970; Abhyankar \& Fymat 1971). The consideration of diffuse reflection of the light beam from semi-infinite atmosphere and the Milne problem, which take into account parameter $\overline{b}_2$, is presented in paper Silant'ev et al. 2014. These problems were considered using the principle of invariance. The problems with sources of nonpolarized radiation is impossible consider in such a manner. We consider these problems using the Green matrix approach.
Below we present of theorethical part of this consideration.

\section{Radiative transfer equation for intensities $I_l(\tau,\mu)$ and $I_r(\tau,\mu)$}

Introducing the (column) vector ${\bf I}$ with the components ($I_l, I_r$) we obtain the following matrix
transfer equation for multiple scattering of continuum radiation on small anisotropic particles
(see Chandrasekhar 1960; Dolginov et al. 1995):
\begin{equation}
\mu \frac{d{\bf I}(\tau,\mu)}{d\tau}= {\bf I}(\tau,\mu)-\overline{b}_1\hat M(\mu^2){\bf K}(\tau)
-\overline{b}_2 K_0(\tau)\left (\begin {array}{c} 1 \\ 1 \end{array}\right )
-\frac{q}{2}Q(\tau)\left (\begin {array}{c} 1 \\ 1 \end{array} \right ),
\label{1}
\end{equation}

\noindent where $q=N_a\sigma_a/(N_s\sigma_s+N_a\sigma_a)$ is the probability of light absorption, $N_a$ and $N_s$ are
the number densities of absorbing (the grains) and scattering particles, respectively;
$d\tau=(N_s\sigma_s+N_a\sigma_a)dz$ determines the dimensionless optical depth, $\sigma_s$ and $\sigma_a$ are the
cross sections of scattering and absorption.

The scattering cross section of small particles (dust grains, molecules) is
$\sigma_s=(8\pi/3)(\omega/c)^4(b_1+3b_2)$, where $\omega=2\pi\nu$ is cyclic frequency of light, $c$ is the speed of
light. For freely (chaotically) oriented particles $\sigma_s$ is independent of the polarization of incident electromagnetic
wave ${\bf E}(\omega)$. The values $b_1$ and $b_2$ are related to polarizability tensor $\beta_{ij}(\omega)$ of
a particle as a whole. Induced dipole moment of a particle, as whole, is equal to
$p_i(\omega)=\beta_{ij}(\omega)E_j(\omega)$. Anisotropic particle with axial symmetry is characterized by two
polarizabilities - along the symmetry axis $\beta_{\parallel}(\omega)$, and in transverse direction
$\beta_{\perp}(\omega)$. For such particles

\[
b_1=\frac{1}{9}|2\beta_{\perp}+\beta_{\parallel}|^2+\frac{1}{45}|\beta_{\parallel}-\beta_{\perp}|^2,
\]
\begin{equation}
b_2=\frac{1}{15}|\beta_{\parallel}-\beta_{\perp}|^2.
\label{2}
\end{equation}
\noindent In transfer equation we use the dimensionless parameters
\begin{equation}
\overline{b}_1=\frac{b_1}{b_1+3b_2},\,\,\, \overline{b}_2=\frac{b_2}{b_1+3b_2},\,\,\,\overline{b}_1+3\overline{b}_2=1.
\label{3}
\end{equation}
\noindent For needle like particles ($|\beta_{\parallel}|\gg |\beta_{\perp}|)$ parameters $\overline{b}_1=0.4,\,
\overline{b}_2=0.2$, and for plate like particles ($|\beta_{\perp}|\gg |\beta_{\parallel}|$) we have
$\overline{b}_1=0.7,\, \overline{b}_2=0.1$.

Parameter $\overline{b}_2$ describes the depolarization of radiation, scattered by freely oriented anisotropic
particles.
The matrix $\hat M(\mu^2)$ has the form (see Lenoble 1970; Abhyankar and Fymat 1971):
\begin{equation}
\hat M(\mu^2)=\frac{\sqrt{3}}{2}\left (\begin{array}{ll}\mu^2\,\,,\,(1-\mu^2)\sqrt{2}\,\\ \,1\,\,\,\,,\,0\,
\end{array}\right )\,\,\,.
\label{4}
\end{equation}
\noindent Note that the superscript T will be used for matrix transpose.

The vector ${\bf K}(\tau)$ and scalar $K_0(\tau)$ are determined by the expressions:

\begin{equation}
{\bf K}(\tau)=\frac{1-q}{2}\int\limits_{-1}^1\,d\mu\hat M^T(\mu^2){\bf I}(\tau,\mu),
\label{5}
\end{equation}

\begin{equation}
K_0(\tau)=\frac{3(1-q)}{4}\int\limits_{-1}^1\,d\mu(\, I_l(\tau,\mu)+I_r(\tau,\mu)).
\label{6}
\end{equation}
\noindent It is easy seen from these expressions that

\begin{equation}
K_0(\tau) =\frac{\sqrt{3}}{2}(2K_l(\tau)+\sqrt{2}K_r(\tau)).
\label{7}
\end{equation}

Usual source of radiation is the thermal radiation
\begin{equation}
Q(\tau)=B_{\omega}(T(\tau)),
\label{8}
\end{equation}
\noindent where $B_{\omega}(T)$ is the Planck function with the temperature, depending on the optical depth $\tau$. This
dependence appears as a result of consideration of various models for atmospheres.
The main features of our investigation are consideration of problems with depolarization parameter
$\overline{b}_2$, and
taking into account the true absorbtion (parameter $q$) (i.e., the existence of absorbing grains in an atmosphere.)

\section{The integral equation for ${\bf K}(\tau)$}

The radiative transfer equation (1) can be written in the form:
\begin{equation}
\mu \frac{d{\bf I}(\tau,\mu)}{d\tau}= {\bf I}(\tau,\mu)-{\bf S}(\tau,\mu^2),
\label{9}
\end{equation}
\noindent where the vector ${\bf S}(\tau,\mu^2)$ is described by the expression:
\begin{equation}
{\bf S}(\tau,\mu^2)=\hat A(\mu^2){\bf K}(\tau)+
\frac{q}{2}Q(\tau)\left (\begin{array}{c}1\\1 \end{array}\right).
\label{10}
\end{equation}
\noindent Here the matrix $\hat A(\mu^2)$ appears from the relation:
\begin{equation}
\hat A(\mu^2){\bf K}(\tau)\equiv \overline {b}_1\hat M(\mu^2){\bf K}(\tau)+\overline {b}_2\frac{\sqrt{3}}{2}(2K_l(\tau)+\sqrt{2}K_r(\tau))\left (\begin{array}{c}1\\1 \end{array}\right).
\label{11}
\end{equation}
\noindent The explicit form of matrix $\hat A(\mu^2)$ is the following:
\begin{equation}
\hat A(\mu^2)=\frac{\sqrt{3}}{2}\left (\begin{array}{ll}\overline{b}_1\mu^2+2\overline{b}_2\,\,,\,(\overline{b}_1(1-\mu^2)+\overline{b}_2)\sqrt{2}\,\\ \,\overline{b}_1+2\overline{b}_2\,\,\,\,,\,\sqrt{2}\,\,\overline{b}_2\,
\end{array}\right )\,.
\label{12}
\end{equation}
The vector ${\bf I}(\tau,\mu)$, according to Eq.(9), can be written as:
\[
{\bf I}(\tau,\mu)=\int\limits_{\tau}^{\infty}\frac{d\tau'}{\mu}
\exp{\left(-\frac{|\tau-\tau'|}{\mu}\right)}\,{\bf S}(\tau',\mu^2),\,\,\,\mu \ge 0,
\]
\begin{equation}
{\bf I}(\tau,\mu)=\int\limits_{0}^{\tau}\frac{d\tau'}{|\mu |}
\exp{\left (-\frac{|\tau-\tau'|}{|\mu |}\right)}{\bf S}(\tau',\mu^2),\,\,\,\mu \le 0.
\label{13}
\end{equation}

Substitution these expressions into Eq.(5), and taking into account that matrix $\hat M^{T}$ and vector ${\bf S}$ are even
functions of $\mu$, we derive the expression:
\begin{equation}
{\bf K}(\tau)=\frac{1-q}{2}\int\limits_0^{\infty}d\tau'\int\limits_0^1\frac{d\mu}{\mu}\hat M^{T}(\mu^2)\,{\bf S}(\tau',\mu^2)\exp{\left(-\frac{|\tau-\tau'|}{\mu}\right)}.
\label{14}
\end{equation}
\noindent Using the formulas (10) and (13), we obtain the integral equation for ${\bf K}(\tau)$:
\begin{equation}
{\bf K}(\tau)={\bf g}(\tau)+\int\limits_0^\infty d\tau'\hat L(|\tau-\tau'|){\bf K}(\tau').
\label{15}
\end{equation}
\noindent The matrix kernel of this integral equation is the following:
\begin{equation}
\hat L(|\tau-\tau'|)=\frac{3(1-q)}{8}\int\limits_0^1\frac{d\mu}{\mu}\exp{\left(-\frac{|\tau-\tau'|}{\mu}\right)}\,\hat a(\mu^2).
\label{16}
\end{equation}
\noindent Here $\hat M^{T}\hat A=3/4\hat a(\mu^2)$. The explicit form of matrix $\hat a(\mu^2)$ is the following:
\begin{equation}
\hat a(\mu^2)=\left (\begin{array}{ll}\overline{b}_1(1+\mu^4)+2\overline{b}_2 (1+\mu^2)\,\,,\,\sqrt{2}(\overline{b}_1\mu^2(1-\mu^2)+\overline{b}_2(1-\mu^2))\,\\ \,\,\sqrt{2}\,\overline{b}_1\mu^2(1-\mu^2)+2\sqrt{2}\,\overline{b}_2(1-\mu^2)\,\,\,\,,\,2\overline{b}_1(1-\mu^2)^2+2\overline{b}_2(1-\mu^2)\,
\end{array}\right )\,\,\,.
\label{17}
\end{equation}

The free term ${\bf g}(\tau)$ has the form:
\begin{equation}
{\bf g}(\tau)=\frac{q(1-q)\sqrt{3}}{8}\int\limits_0^{\infty}d\tau'\int\limits_0^1\frac{d\mu}{\mu}Q(\tau')\exp{\left(-\frac{|\tau-\tau'|}{\mu}\right)}\left (\begin{array}{c}(1+\mu^2)\\ \sqrt{2}(1-\mu^2) \end{array}\right).
\label{18}
\end{equation}

\section{Solution of integral equation (15) using the Green matrix}

According to the standard theory of integral equations (see, for example, Tricomi 1957), the solution of equation (15) can be presented in the form:
\begin{equation}
{\bf K}(\tau)={\bf g}(\tau)+\int\limits_0^{\infty}d\tau'\hat  R(\tau,\tau'){\bf g}(\tau'),
\label{19}
\end{equation}
\noindent where the Green matrix $\hat R(\tau,\tau')$ obeys
the integral equation:
\begin{equation}
\hat R(\tau,\tau')=\hat L(|\tau-\tau'|)+\int\limits_0^{\infty}d\tau''\hat L(|\tau-\tau''|)\hat R(\tau'',\tau').
\label{20}
\end{equation}
\noindent It follows from this equation that
\begin{equation}
\hat R(\tau,0)=\hat L(\tau)+\int\limits_0^{\infty}d\tau'\hat L(|\tau-\tau'|)\hat R(\tau',0).
\label{21}
\end{equation}
\noindent This equation coincides formally with Eq.(15), and, consequently, $\hat R(\tau,0)$ can be written in the form:
\[
\hat R(\tau,0)=\hat L(\tau)+\int\limits_0^{\infty}d\tau'\hat R(\tau,\tau')\hat L(\tau')\equiv
\]
\begin{equation}
\frac{3(1-q)}{8}\int\limits_0^1\frac{d\mu}{\mu}\,\left[\exp{
\left(-\frac{\tau}{\mu}\right)}+\tilde {\hat R}(\tau,\frac{1}{\mu})\right]\,\hat a(\mu^2),
\label{22}
\end{equation}
\noindent where $\tilde {\hat R}(a)$ denotes the Laplace transform:
\begin{equation}
\tilde {\hat R}(a)=\int\limits_0^{\infty}d\tau \exp{(-a\tau)}\hat R(\tau).
\label{23}
\end{equation}
As is known, the Green matrix  $\hat R(\tau,\tau')$ also obeys the another equation:
\begin{equation}
\hat R(\tau,\tau')=\hat L(|\tau-\tau'|)+\int\limits_0^{\infty}d\tau''\hat R(\tau,\tau'')\hat L(|\tau''-\tau'|).
\label{24}
\end{equation}
\noindent From this equation we obtain
\begin{equation}
\hat R(0,\tau )=\hat L(\tau )+\int\limits_0^{\infty}d\tau'\hat R(0,\tau')\hat L(|\tau'-\tau|).
\label{25}
\end{equation}
\noindent Analogously to equation (22), the $\hat R(0,\tau)$
can be written in the form:
\[
\hat R(0,\tau)=\hat L(\tau)+\int\limits_0^{\infty}d\tau'\hat L(\tau')\hat R(\tau',\tau)\equiv
\]
\begin{equation}
\frac{3(1-q)}{8}\int\limits_0^1\frac{d\mu}{\mu}\,\hat a(\mu^2)
\left[\exp{\left(-\frac{\tau}{\mu}\right)}+\tilde {\hat R}(\frac{1}{\mu},\tau)\right].
\label{26}
\end{equation}

 Differentiation of Eq.(20) (or (24)) gives rise to relation: 
\begin{equation}
\frac{\partial \hat {R}(\tau,\tau')}{\partial\tau}+
\frac{\partial \hat {R}(\tau,\tau')}{\partial\tau'}=
\hat R(\tau,0)\hat R(0,\tau').
\label{27}
\end{equation}
\noindent This relation means that the total Green matrix $\hat {R}(\tau,\tau')$ can be obtained, if we know two auxiliary matrices, depending on one variable - $\tau$ or $\tau'$.
The Laplace transform over $\tau$ and over $\tau'$ of the matrix $\hat R(\tau,\tau')$ we denote as
\begin{equation}
\tilde{\tilde {\hat R}}(a,b)=\int\limits_0^{\infty}d\tau\,\int\limits_0^{\infty}d\tau'\exp{(-a\tau)}\exp{(-b\tau')}\hat R(\tau,\tau').
\label{28}
\end{equation}

Taking this double Laplace transform in equation (27), we obtain the relation:
\begin{equation}
\tilde{\tilde {\hat R}}(a,b)=\frac{1}{a+b}[\,\tilde{\hat R}(a,0)+\tilde{\hat R}(0,b)+\tilde{\hat R}(a,0)\tilde{\hat R}(0,b)].
\label{29}
\end{equation}
\noindent   
Taking the Laplace transform over variable $\tau$ in  equations (22) and (26), and using the relation (29), we obtain the system equations for $\tilde {\hat R}(a,0)$ and
$\tilde{\hat R}(0,a)$.

 It is useful to introduce the new matrices
\[
\hat {H}^{(1)}(\frac{1}{b})=\hat {I}+\tilde{\hat R}(0,b),
\]
\begin{equation}
\hat {H}^{(2)}(\frac{1}{b})=\hat {I}+\tilde {\hat R}(b,0).
\label{30}
\end{equation}
\noindent For these new matrices we obtain the following system of equations:
\[
\hat {H}^{(1)}(\frac{1}{b})=\hat {I}+\frac{3(1-q)}{8}\int\limits_0^1d\mu\frac{\hat {a}(\mu^2)}{b\mu+1}\hat {H}^{(2)}(\mu)\hat {H}^{(1)}(\frac{1}{b}),
\]
\begin{equation}\hat {H}^{(2)}(\frac{1}{b})=\hat {I}+\frac{3(1-q)}{8}\int\limits_0^1d\mu\frac{1}{b\mu+1}\hat {H}^{(2)}(\frac{1}{b})\hat {H}^{(1)}(\mu)\hat {a}(\mu^2).
\label{31}
\end{equation}

 For the sake of convenience, we introduce the notation $1/b =z$. Then equations (31) acquire more simple form:
\[
\hat {H}^{(1)}(z)=\hat {I}+\frac{3(1-q)}{8}z\int\limits_0^1\,d\mu\frac{\hat {a}(\mu^2)}{z+\mu}\hat {H}^{(2)}(\mu)\hat {H}^{(1)}(z),
\]
\begin{equation}
\hat {H}^{(2)}(z)=\hat {I}+\frac{3(1-q)}{8}z\int\limits_0^1\,d\mu\frac{1}{z+\mu}\hat {H}^{(2)}(z)\hat {H}^{(1)}(\mu)\hat {a}(\mu^2).
\label{32}
\end{equation}

 It is seen, that to know $\hat {H}^{(1)}(z)$ and
$\hat {H}^{(2)}(z)$ we are to know $\hat {H}^{(1)}(\mu)$ and
$\hat {H}^{(2)}(\mu)$. Taking $z=\mu$, we derive from the system  (32) the separate system of equations for $\hat {H}^{(1)}(\mu)$ and $\hat {H}^{(2)}(\mu)$:
\[
\hat {H}^{(1)}(\mu)=\hat {I}+\frac{3(1-q)}{8}\mu\int\limits_0^1\,d\mu'\frac{\hat {a}(\mu'^2)}{\mu+\mu'}\hat {H}^{(2)}(\mu')\hat {H}^{(1)}(\mu),
\]
\begin{equation}
\hat {H}^{(2)}(\mu)=\hat {I}+\frac{3(1-q)}{8}\mu\int\limits_0^1\,d\mu'\frac{1}{\mu+\mu'}\hat {H}^{(2)}(\mu)\hat {H}^{(1)}(\mu')\hat {a}(\mu'^2).
\label{33}
\end{equation}
\noindent Thus, the system (33) plays the main role in the Green matrix approach. Note that for the case $\overline {b}_2=0$ the matrix ${\hat {a}(\mu^2)}={\hat {a}(\mu^2)}^T$. In this case we have $\hat {H}^{(1)}(\mu)=\hat {H}^{(2)}(\mu)^T$. 
  
The system (33) is nonlinear system of matrix equations.   It is known (see Sobolev 1969) that scalar $ H(\mu)$-function also obeys the linear equation. Let us derive the
linear equations for our matrix functions $\hat {H}^{(1)}(\mu)$ and $\hat {H}^{(2)}(\mu)$.
\section{Linear equations for matrix functions $\hat {H}^{(1)} $ and $\hat {H}^{(2)}$}
Let us multiply equation for $H^{(1)}(\mu)$ by the value $\lambda z\hat a(\mu^2)/(z-\mu)$ from the right side of equation and take the integration over $\mu$ in the interval (0,1). For brevity, we introduced $\lambda=\frac{3(1-q)}{8}$. As a result, we obtain the expression:
\begin{equation}
\lambda z\int\limits_0^1\,d\mu\frac{\hat {H}^{(1)}(\mu)\hat {a}(\mu^2)}{z-\mu}=\lambda z\int\limits_0^1\,d\mu\,\frac{\hat {a}(\mu^2)}{z-\mu}+\lambda^2 z\int\limits_0^1\,d\mu\int\limits_0^1\,d\mu'\frac{\hat {a}(\mu'^2)\mu}{(\mu+\mu')(z-\mu)}\hat {H}^{(2)}(\mu')\hat {H}^{(1)}(\mu)\hat {a}(\mu^2).
\label{34}
\end{equation} 
\noindent The same operation with $(z+\mu)$ gives:
\begin{equation}
\lambda z\int\limits_0^1\,d\mu\frac{\hat {H}^{(1)}(\mu)\hat {a}(\mu^2)}{z+\mu}=\lambda z\int\limits_0^1\,d\mu\,\frac{\hat {a}(\mu^2)}{z+\mu}+\lambda^2 z\int\limits_0^1\,d\mu\int\limits_0^1\,d\mu'\frac{\hat {a}(\mu'^2)\mu}{(\mu+\mu')(z+\mu)}\hat {H}^{(2)}(\mu')\hat {H}^{(1)}(\mu)\hat {a}(\mu^2).
\label{35}
\end{equation}
 
Now let us multiply equation for $\hat {H}^{(2)}(\mu)$ by the value $\lambda z\hat a(\mu^2)/(z-\mu)$ from the left side of equation and take the usual integration over $\mu$ in the interval (0,1). As a result, we obtain the expression:
\begin{equation}
\lambda z\int\limits_0^1\,d\mu\,\frac{\hat {a}(\mu^2)\hat {H}^{(2)}(\mu)}{z-\mu}=\lambda z\int\limits_0^1\,d\mu\,\frac{\hat {a}(\mu^2)}{z-\mu}+\lambda^2 z\int\limits_0^1\,d\mu\int\limits_0^1\,d\mu'\frac{\hat {a}(\mu^2)\mu}{(\mu+\mu')(z-\mu)}\hat {H}^{(2)}(\mu)\hat {H}^{(1)}(\mu')\hat {a}(\mu'^2).
\label{36}
\end{equation} 
\noindent The same operation with $(z+\mu)$ gives:
\begin{equation}
\lambda z\int\limits_0^1\,d\mu\,\frac{\hat {a}(\mu^2)\hat {H}^{(2)}(\mu)}{z+\mu}=\lambda z\int\limits_0^1\,d\mu\,\frac{\hat {a}(\mu^2)}{z+\mu}+\lambda^2 z\int\limits_0^1\,d\mu\int\limits_0^1\,d\mu'\frac{\hat {a}(\mu^2)\mu}{(\mu+\mu')(z+\mu)}\hat {H}^{(2)}(\mu)\hat {H}^{(1)}(\mu')\hat {a}(\mu'^2).
\label{37}
\end{equation}
The summation of Eqs.(34) and (37) gives rise:
\[
\lambda z\int\limits_0^1\,d\mu\frac{\hat {H}^{(1)}(\mu)\hat {a}(\mu^2)}{z-\mu}+\lambda z
\int\limits_0^1\,d\mu\frac{\hat {a}(\mu^2)\hat {H}^{(2)}(\mu)}{z+\mu}=\lambda z\int\limits_0^1\,d\mu
\frac{\hat {a}(\mu^2)}{z-\mu}+\lambda z\int\limits_0^1d\mu\frac{\hat {a}(\mu^2)}{z+\mu}+
\]
\[
\lambda^2 z\int\limits_0^1\,d\mu\int\limits_0^1\,d\mu'\frac{\hat {a}(\mu'^2)\mu}{(\mu+\mu')(z-\mu)}
\hat {H}^{(2)}(\mu')\hat {H}^{(1)}(\mu)\hat {a}(\mu^2)+
\]
\begin{equation}
\lambda^2 z\int\limits_0^1d\mu\int\limits_0^1d\mu'\frac{\hat {a}(\mu^2)\mu}{(\mu+\mu')(z+\mu)}
\hat {H}^{(2)}(\mu)\hat {H}^{(1)}(\mu')\hat {a}(\mu'^2).
\label{38}
\end{equation}
In last double integral we take the substitutions $\mu\to \mu'$ and $\mu'\to \mu$. After that the sum of two last integrals in Eq.(38) acquires the form:
\begin{equation}
\lambda^2 z\int\limits_0^1d\mu\int\limits_0^1d\mu'\frac{\hat {a}(\mu'^2)}{(z+\mu')(z-\mu)}\hat {H}^{(2)}(\mu')\hat {H}^{(1)}(\mu)\hat {a}(\mu^2).
\label{39}
\end{equation}
From first equation in system (32) it follows that
\begin{equation}
\lambda z\int\limits_0^1d\mu\frac{\hat {a}(\mu^2)\hat {H}^{(2)}(\mu)}{z+\mu}=\hat {I}-(\hat {H}^{(2)}(z))^{-1}.
\label{40}
\end{equation}

 This integration exists in Eq.(39) and in left part of Eq.(38). Using the relation (40) in Eqs.(38) and (39), we
obtain the following linear equation:
\begin{equation}
\hat {H}^{(1)}(z)\left (\hat {I}-2\lambda z^2\int\limits_0^1\,d\mu\frac{\hat {a}(\mu^2)}{z^2-\mu^2}\right )=
\hat {I}-\lambda z\int\limits_0^1\,d\mu\frac{\hat {H}^{(1)}(\mu)\hat {a}(\mu^2)}{z-\mu}.
\label{41}
\end{equation}

 The summation of equations (35) and (36) and using the relation
\begin{equation}
\lambda z\int\limits_0^1\,d\mu\frac{\hat {H}^{(1)}(\mu)\hat {a}(\mu^2)}{z+\mu}=\hat {I}-(\hat {H}^{(1)}(z))^{-1},
\label{42}
\end{equation}
\noindent which follows from the second equation in system (32), we obtain the following linear equation for $\hat {H}^{(2)}(z)$:
\begin{equation}
\left (\hat {I}-2\lambda z^2\int\limits_0^1\,d\mu\frac{\hat {a}(\mu^2)}{z^2-\mu^2}\right )\hat {H}^{(2)}(z)=
\hat {I}-\lambda z\int\limits_0^1\,d\mu\frac{\hat {a}(\mu^2)\hat {H}^{(2)}(\mu)}{z-\mu}.
\label{43}
\end{equation}

 Equations (41) and (43) are the equations with singular kernels. The solution such scalar equations is presented in the book (Gakhov 1977).

 Some times the expression
\begin{equation}
\int\limits_0^1\,d\mu\frac{\hat {H}^{(1)}(\mu)}{z-\mu}
\label{44}
\end{equation}
\noindent is used. To derive this expression, we represent formula (17) in the form:
\begin{equation}
\hat {a}(\mu^2)=\hat {a}_0 +\hat {a}_2\mu^2 +\hat {a}_4\mu^4,
\label{45}
\end{equation}
\noindent where the matrices $\hat {a}_0,\hat {a}_2$ and $\hat {a}_4$ are independent of $\mu$:
\[
\hat {a}_0=\left (\begin{array}{ll}\overline{b}_1+2\overline{b}_2\,\,,\,\overline{b}_2\sqrt{2}\\ \,\,2\overline{b}_2\sqrt{2}\,\,\,\,,\,2\overline{b}_1+2\overline{b}_2\,
\end{array}\right )\,\,\,,
\]
\[
\hat {a}_2=\left (\begin{array}{ll}2\overline{b}_2 \,\,,\,(\overline{b}_1-\overline{b}_2)\sqrt{2}\\ \,\,\sqrt{2}(\overline{b}_1+2\overline{b}_2)\,\,\,\,,\,-4\overline{b}_1-2\overline{b}_2\sqrt{2}
\end{array}\right )\,\,\,,
\]
\begin{equation}
\hat {a}_4=\left (\begin{array}{ll}\overline{b}_1\,\,,\,-\overline{b}_1\sqrt{2}\,\\ \,\,-\overline{b}_1\sqrt{2}\,\,\,\,,\,2\overline{b}_1\,
\end{array}\right )\,\,\,.
\label{46}
\end{equation}
Using these relations, we obtain the following expression:
\[
\lambda\,z\int\limits_0^1\,d\mu\frac{\hat {H}^{(1)}(\mu)}{z-\mu}=\hat {H}^{(1)}(z)\left (\hat {I}+2\lambda z^2\int\limits_0^1\,d\mu\frac{\hat {a}(\mu^2)}{\mu^2-z^2}\right ) \hat {a}(z^2)^{-1}-\hat {a}(z^2)^{-1}-
\]
\begin{equation}
\lambda z\left[(\hat {H}^{(1)}_1+z\hat {H}^{(1)}_0)\hat {a}_2+
(\hat {H}^{(1)}_3+z\hat {H}^{(1)}_2+z^2\hat {H}^{(1)}_0))\hat  {a}_4\right ]\hat {a}(z^2)^{-1}.
\label{47}
\end{equation}
\noindent Here $\hat {H}^{(1)}_n$ denote the moments of
the matrix $\hat {H}^{(1)}$:
\begin{equation}
\hat {H}^{(1)}_n=\int\limits_0^1\,d\mu\,\mu^{n}\hat {H}^{(1)}(\mu).
\label{48}
\end{equation}
Remember that inverse 2X2 matrix  has very simple form:
\begin{equation}
\hat {B}^{-1}=\left (\begin{array}{ll}B_1\,\,,\,B_2\,\\ \,B_3\,\,\,\,,\,B_4\,\,\,\,
\end{array}\right )^{-1}=\frac{1}{B_1B_4-B_2B_3}\left (\begin{array}{ll}B_4\,\,,\,-B_2\,\\ \,-B_3\,\,\,\,,\,B_1\,\end{array}\right )\,\,\,.
\label{49}
\end{equation} 

\section{Formulas for outgoing radiation}

According to Eq. (9) and (10), the vector ${\bf I}(0,\mu)$, describing the outgoing radiation, has the form:
\begin{equation}
{\bf I}(0,\mu)=\int\limits_0^{\infty}\frac{d\tau }{\mu}\exp{\left(-\frac{\tau}{\mu}\right)}{\bf S}(\tau,\mu)\equiv \int\limits_0^{\infty}\frac{d\tau }{\mu}\exp{\left(-\frac{\tau}{\mu}\right)}\left (\hat {A}(\mu^2){\bf K}(\tau)+\frac{q}{2}Q(\tau))\left (\begin{array}{c}1\\1 \end{array}\right)\right ),
\label{50}
\end{equation}
\noindent i.e. this expression proportional to the Laplace transform over variable $\tau/\mu $.

 The vector ${\bf K}(\tau)$ is presented in Eq.(19), where the
vector ${\bf g}(\tau)$ is given in Eq.(18). The Laplace transform of vector ${\bf K}(\tau)$ can be written in the form:
\begin{equation}
\tilde{{\bf K}}(\frac{1}{\mu})=\int\limits_0^{\infty}d\tau \exp{\left(-\frac{\tau}{\mu}\right)}{\bf K}(\tau)\equiv \int\limits_0^{\infty}d\,\tau\left ( \exp{\left(-\frac{\tau}{\mu}\right)}\,\hat {I}+\tilde{\hat R}(\frac{1}{\mu},\tau){\bf g}(\tau)\right ).
\label{51}
\end{equation}
 Thus , expression (50) acquires the form:
\begin{equation}
{\bf I}(0,\mu)=\frac{1}{\mu}\left \{ \hat A(\mu^2)\left [
\tilde {\bf g}(\frac{1}{\mu}) + \int\limits_0^{\infty}d\tau 
\tilde {\hat {R}}(\frac{1}{\mu},\tau){\bf g}(\tau)\right]
+\frac{q}{2}\tilde {Q}(\frac{1}{\mu})\left (\begin{array}{c}1\\1 \end{array}\right) \right\}.
\label{52}
\end{equation}

 Note that the value $Q(\tau)$ characterises the distribution of sources of nonpolarized radiation in an atmosphere.
For source function $Q_h(\tau)=Q_h\exp{(-h\tau)}$ the expression for ${\bf I}(0,\mu)$ acquires comparatively simple form. In this case the expression depends on $\hat {H}^{(1)}(\mu)$ and 
$\hat {H}^{(2)}(\mu)$, not on matrix $\hat {R}(\tau,\tau')$.

The sources of types $Q_n(\tau )=Q_n\tau^n$ are related with exponential source by the simple formula:
\begin{equation}
Q_n(\tau)=(-1)^n Q_n \frac{d^n}{dh^n}\exp{(-h\tau)}|_{h=0} .
\label{53}
\end{equation}
It means that the sources of type
\begin{equation}
Q(\tau)=Q_h\exp{(-h\tau)}+Q_0 +Q_1\tau+Q_2\tau^2 +...
\label{54}
\end{equation}
\noindent can be considered  on the base of exponential source.

Consider this case in detail. Taking $Q_h(\tau)=Q_h\exp{(-h\tau)}$  in Eq.(18) we obtain
\begin{equation}
{\bf g}(\tau)=\frac{\sqrt{3}}{8}q(1-q)Q_h\int\limits_0^1\frac{d\mu}{\mu}\left (\begin{array}{c}(1+\mu^2)\\ \sqrt{2}(1-\mu^2) \end{array}\right)\left [\frac{\exp{\left(-\frac{\tau}{\mu}\right)}-\exp{\left(-h\tau\right)}}{h-\frac{1}{\mu}}+\frac{\exp{\left(-h\tau\right)}}{h+\frac{1}{\mu}}\right].
\label{55}
\end{equation}
Substituting this expression into Eq. (52) and taking into account formula (29) for double Laplace transform of matrix
$\hat {R}(\tau,\tau')$, we derive the following expression:
\[
{\bf I}(0,\mu)=Q_h\left \{\frac{q}{2}\frac{1}{h\mu+1}\left (\begin{array}{c}1\\1 \end{array}\right)+
\frac{\sqrt{3}}{8}q(1-q)
\hat {A}(\mu^2)\hat {H}^{(2)}(\mu)\int\limits_0^1\,d\mu'
\left [\frac{\hat {H}^{(1)}(\mu')\mu'}{(h\mu'-1)(\mu+\mu')}-\right.
\right.
\]
\begin{equation}
\left.\left.\frac{\hat {H}^{(1)}(\frac{1}{h})}{(h\mu'-1)(h\mu+1)}
+\frac{\hat {H}^{(1)}(\frac{1}{h})}{(h\mu'+1)(h\mu+1)}\right ]
\left (\begin{array}{c}(1+\mu'^2)\\ \sqrt{2}(1-\mu'^2) \end{array}\right)\right \}.
\label{56}
\end{equation} 
For homogeneous source $Q_0$ we are take $h=0$. In this case we
obtain from Eq. (56):
\[
{\bf I}(0,\mu)=Q_0\left \{\frac{q}{2}\left (\begin{array}{c}1\\1 \end{array}\right)+
\frac{\sqrt{3}}{8}q(1-q)
\hat {A}(\mu^2)\hat {H}^{(2)}(\mu)\int\limits_0^1\,d\mu'
\left[-\frac{\hat {H}^{(1)}(\mu')\mu'}{(\mu+\mu')}+\right.
\right.
\]
\begin{equation}
\left.\left.2\hat {H}^{(1)}(\infty)
\right ]
\left (\begin{array}{c}(1+\mu'^2)\\ \sqrt{2}(1-\mu'^2) \end{array}\right)\right\}. 
\label{57}
\end{equation}
\noindent Taking $z=\infty$ in first equation of system (32), we obtain:
\begin{equation}
\hat {H}^{(1)}(\infty)=\left [\hat {I}-\frac{3(1-q)}{8}\left (
\hat {a}_0\hat {H}^{(2)}_0+\hat {a}_2\hat {H}^{(2)}_2+
\hat  {a}_4\hat {H}^{(2)}_4\right)\right]^{-1}.
\label{58}
\end{equation} 

 The matrices $\hat {a}_0$, $\hat {a}_2$ and $\hat {a}_4$ are presented in Eq. (46). Definition of moments  $\hat {H}^{(2)}_n$ is given 
in Eq. (48). The outgoing radiation for the source $Q_1\tau$ can be derived according to Eq.(53):
\begin{equation}
{\bf I}(0,\mu)=Q_1\left \{\frac{q\mu}{2}\left (\begin{array}{c}1\\1 \end{array}\right)+
\frac{\sqrt{3}}{8}q(1-q)
\hat {A}(\mu^2)\hat {H}^{(2)}(\mu)\int\limits_0^1\,d\mu'
\frac{\hat {H}^{(1)}(\mu')\mu'^2}{(\mu+\mu')}
\left (\begin{array}{c}(1+\mu'^2)\\ \sqrt{2}(1-\mu'^2) \end{array}\right)\right\}. 
\label{59}
\end{equation}

Analogously can be derived formulas for  outgoing radiation for sources of type $\sim \tau^2, \tau^3$ etc.

\section{Conclusion}

In this paper we consider the role of anisotropy of molecules and dust grains in the process of depolarization
of multiple scattered light in semi-infinite plane-parallel atmospheres with the  sources of unpolarized 
radiation. The consideration uses the matrix Green function
$\hat {H}(\tau,\tau')$  approach. This matrix is connected with two auxiliar matrices,
 depending on one variable $\tau$ or $\tau'$,  which obey the system of nonlinear equations similar to known system of equations for scalar $H(\mu)$- functions of
Chandrasekhar. It is shown that for sources of type $\exp{(-h\tau)}$ and sources of type $ \tau^0, \tau, \tau^2 $ etc. the
formulas for outgoing radiation depend on matrix $H$-functions and have rather simple form. 

This research was supported by the Program of Prezidium of RAS No21,
the Program of the Department of Physical Sciences of RAS No17, the Federal
Target Program ''Science and Scientific-Pedagogical Personnel of Innovative Russia'' XXXVII in turn
- the action 1.2.1,
and by the Grant from President of the Russian Federation ''The Basic Scientific Schools'' (NSh-1625.2012.2).

\end{document}